\documentclass[prl,aps,twocolumn,showpacs,amsmath,amssymb]{revtex4-1}

\usepackage{graphicx}
\usepackage{psfrag}
\usepackage{epsfig}
\usepackage{color}
\usepackage[tight]{subfigure}
\definecolor{dred}{rgb}{0.7,0.0,0.0}
\usepackage{dcolumn}% Align table columns on decimal point
\usepackage{bm}% bold math
\allowdisplaybreaks 
\usepackage{braket}
\usepackage{ulem} % to strike things out
\usepackage{hyperref}
\hypersetup{colorlinks=true, linkcolor=blue, citecolor=blue, filecolor=blue, urlcolor=blue, pdftitle=, pdfauthor= 	, pdfsubject=, pdfkeywords=}

\newcommand{\bs}[1]{\ensuremath{\boldsymbol{#1}}}

\definecolor{orange}{rgb}{1,0.5,0}
\definecolor{black}{rgb}{0,0,0}

\newcounter{defcounter}
\newenvironment{sequation}{%
\refstepcounter{defcounter}

\begin{equation}}
{\end{equation}}

\begin{document}

\title{
Shiba chains of scalar impurities on unconventional superconductors 
}

\date{\today}

\author{Titus Neupert} 
\affiliation{
Princeton Center for Theoretical Science,
Princeton University,
Princeton, NJ 08544, USA
            }

\author{A. Yazdani} 
\affiliation{
Department of Physics,
Princeton University,
Princeton, NJ 08544, USA
            }

\author{B. Andrei Bernevig} 
\affiliation{
Department of Physics,
Princeton University,
Princeton, NJ 08544, USA
            }

\begin{abstract}
We show that a chain of \emph{nonmagnetic} impurities 
deposited on a fully gapped two- or three-dimensional superconductor can 
become a topological one-dimensional superconductor 
with protected Majorana bound states at its end. 
A prerequisite is that the pairing potential of the underlying superconductor 
breaks the spin-rotation symmetry, as it is generically the case in  systems with
strong spin-orbit coupling. 
We illustrate this mechanism for a spinless triplet-superconductor ($p_x+ i p_y$) and a 
time-reversal symmetric Rashba superconductor with a mixture of singlet and triplet pairing.
For the latter, we show that the impurity chain can be topologically nontrivial
even if the underlying superconductor is topologically trivial. 
\end{abstract}

\maketitle

Majorana bound states are a distinctive feature of topological superconductors. To verify the exotic properties that theory attributes to them, in particular their non-Abelian braiding statistics,~\cite{Read2000,Ivanov01} Majorana modes have to be obtained as controllable, localized excitations. 
For that reason, significant research efforts are focused on one-dimensional (1D) topological superconductors, in which Majorana states naturally appear as zero-dimensional, i.e., fully localized, end states.~%
\cite{Kitaev01,Lutchyn10,Oreg10, Sau10, Alicea10, Potter10, Alicea11, Halperin12, Stanescu13}

One of the ways to create a topological 1D superconductor is 
through the deposition of a chain of magnetic adatoms on the surface of a thin-film or bulk superconductor.
 If these adatoms order magnetically -- either as a helical magnetic spiral or as a ferromagnet -- the chain can become a 1D topological superconductor, even if the underlying superconductor is of $s$-wave spin-singlet type.~\cite{Choy11,Nadj-Perge13,Nakosai13, Klinovaja13, Braunecker13, Vazifeh13, Pientka13, Po13, Pientka13b,
 Li14,Nadj-Perge14,Pawlak15,Ruby15}
  Ferromagnetic chains of adatoms also require the presence of spin-orbit coupling in the underlying superconductor in order to become topological. The key point in this construction is that magnetic impurities in $s$-wave superconductors feature so-called Shiba midgap bound states,~\cite{Yu65,Shiba68,Rusinov68,Pientka13,Li14,Kaladzhyan15} which can hybridize along the chain and experience band inversion. The appearance of Shiba states can be intuitively understood from ``Anderson's theorem",~\cite{Anderson59} stating that  $s$-wave superconductivity is (locally) suppressed by magnetic impurities.~\cite{Supp}

Here, we exploit a second implication of ``Anderson's theorem", namely the fact that \emph{un}conventional superconducting pairing is suppressed by \emph{non}magnetic impurities.~\cite{Byers93,Choi04,Sau13,Hu13,Kim15} We ask the question under which conditions bound states of a chain of nonmagnetic scalar impurities can hybridize in such a way that they form a 1D topological superconductor.
In particular, such a construction involving nonmagnetic impurities could allow for time-reversal symmetric (TRS) 1D superconductors with Kramers pairs of Majorana end states. 

We consider two examples for the underlying unconventional superconductor: (i) a TRS superconductor with Rashba spin-orbit interaction that mixes $s$-wave singlet and $p$-wave triplet pairing and (ii) a spinless TRS breaking $p$-wave superconductor. 
While the latter is per se a bulk topological superconductor that in itself supports Majorana edge modes, the former can either be in a topological or in a trivial phase, depending on whether the triplet or the singlet pairing dominates. We find that the chain of nonmagnetic impurities can support Majorana end modes, both in the topological and trivial phase of the underlying superconductor, provided that the triplet pair potential is sufficiently large.  Despite the fact that the required fully gapped unconventional or strong-Rashba superconductors are much less abundant than conventional $s$-wave superconductors, several promising examples have been discovered and examined recently. This includes thin-film and interface superconductors such as LaAlO$_3$/SrTiO$_3$~\cite{Reyren07} and single-layer of FeSe on SrTiO$_3$~\cite{Ge14} as well as bulk materials such as Sr$_2$RuO$_4$~\cite{Mackenzie03} and CePt$_3$Si~\cite{Bauer04}. The observation of Majorana bound states at the end of non-magnetic adatom chains deposited on these superconductors would also confirm their unconventional pairing.  

Before we consider a specific model system, let us outline the general derivation of an effective Hamiltonian for the chain of scalar impurities following Ref.~\cite{Pientka13}. We consider a two-dimensional (2D) superconductor with Hamiltonian 
\begin{equation}
H_0:=
\int\frac{\mathrm{d}^2\bs{k}}{(2\pi)^2}
\Psi^\dagger(\bs{k})
\mathcal{H}_{\bs{k}}
\Psi(\bs{k}),
\end{equation}
where $\Psi^\dagger(\bs{k})=(c^\dagger_{\bs{k}\uparrow},c^\dagger_{\bs{k}\downarrow},c_{-\bs{k}\downarrow},-c_{-\bs{k}\uparrow})$ is the 4-spinor-valued  annihilation operator of a Bogoliubov quasiparticle at momentum $\bs{k}$ and $\mathcal{H}_{\bs{k}}$ is the $(4\times4)$ Bogoliubov de-Gennes (BdG) Hamiltonian including the spin and particle-hole grading.~\cite{footnote-3D}

On this superconductor, impurities are deposited at positions $\bs{r}_n= \mathsf{a} n\bs{e}_1,\ n\in\mathbb{Z}$, i.e., with distance $\mathsf{a}$ along the 1-direction. 
For a delta-function density impurity of strength $U$, the impurity Hamiltonian reads
\begin{equation}
\begin{split}
H_{\mathrm{imp}}
=&-\frac{U}{2}\sum_n\Psi^\dagger(\bs{r}_n)X_{03}\Psi(\bs{r}_n)
\\
=&-\frac{U}{2}\int_{-\pi/\mathsf{a}}^{\pi/\mathsf{a}}\frac{\mathrm{d}\mathsf{k}_1}{2\pi}\Phi^\dagger(\mathsf{k}_1)X_{03}\Phi(\mathsf{k}_1),
\end{split}
\label{eq: impurity Hamiltonian}
\end{equation}
with $\mathsf{k}_1\in[-\pi/\mathsf{a},\pi/\mathsf{a})$, 
where we use the notation $X_{\mu\nu}=\sigma_\mu\otimes\tau_\nu$ for the tensor product of Pauli matrices $\sigma_\mu$ and $\tau_\mu$, $\mu=0,\cdots, 3$, that act on spin-space and  on particle-hole space, respectively, with $\tau_0$ and $\sigma_0$ being the identity matrices.
Here, $\Psi(\bs{r})$ is the Fourier transform of $\Psi(\bs{k})$ and the 4-spinor $\Phi^\dagger(\mathsf{k}_1)$ is defined in as
\begin{equation}
\Phi^\dagger(\mathsf{k}_1)
:=
\sum_n \int_{-\infty}^{\infty}\frac{\mathrm{d}k_2}{2\pi}\Psi^\dagger
\left(\mathsf{k}_1+2\pi\,n/\mathsf{a},k_2\right).
\end{equation}
Notice that $\Phi^\dagger(\mathsf{k}_1)$ is periodic in $\mathsf{k}_1$ with period $2\pi/\mathsf{a}$, while the momentum $\bs{k}\in \mathbb{R}^2$. 

The Schroedinger equation for the superconducting electrons in presence of the chain of impurity potentials, governed by the Hamiltonian $H_0+H_{\mathrm{imp}}$,  reads
\begin{equation}
(E_{\mathsf{k}_1}-\mathcal{H}_{\bs{k}})\Psi(\bs{k})
=-\frac{U}{2}\,X_{03}\Phi(\mathsf{k}_1),
\end{equation}
where $\bs{k}=(\mathsf{k}_1+2\pi n/\mathsf{a},k_2)$, $n\in\mathbb{Z}$.
We note that $\mathsf{k}_1$, but not $\bs{k}$, is a good quantum number which can be used to label the energies. 
After multiplication with $(E_{\mathsf{k}_1}-\mathcal{H}_{\bs{k}})^{-1}$ from the left, integration over $k_2$  and summation over all momenta $k_1=\mathsf{k}_1+2\pi\,n/\mathsf{a}$ on both sides, it can be transformed into an eigenvalue equation for $\Phi(\mathsf{k}_1)$
\begin{equation}
\Phi(\mathsf{k}_1)
=\frac{U}{2}R(\mathsf{k}_1,E_{\mathsf{k}_1})
\Phi(\mathsf{k}_1),
\label{eq: self-consistent equation with R}
\end{equation}
where we defined
\begin{equation}
R(\mathsf{k}_1,E_{\mathsf{k}_1}):=
\sum_n \int_{-\infty}^{\infty}\frac{\mathrm{d}k_2}{2\pi}
\left(\mathcal{H}_{\mathsf{k}_1+2\pi n/\mathsf{a},k_2}-E_{\mathsf{k}_1}\right)^{-1}X_{03}
\label{eq: R with the inverse of H}
\end{equation}
as a $(2\pi/\mathsf{a})$-periodic matrix-valued function of ${\mathsf{k}_1}$.
Analytically, the only solution possible is in the case where the impurity bound states hybridize into a band of energies $E_{\mathsf{k}_1}$ well below the bulk gap of the superconductor so that we can perform the expansion
\begin{equation}
R(\mathsf{k}_1,E_{\mathsf{k}_1})= 
g(\mathsf{k}_1)+f(\mathsf{k}_1)E_{\mathsf{k}_1}+\mathcal{O}(E_{\mathsf{k}_1}^2).
\label{eq: R expansion}
\end{equation}
In particular, this expansion is justified as we are primarily interested in topological phase transitions of the Shiba wire at which the energy eigenvalue $E_{\mathsf{k}_1}$ vanishes.
Inserting Eq.~\eqref{eq: R expansion} in Eq.~\eqref{eq: self-consistent equation with R} yields the effective Hamiltonian of the 1D chain of impurity bound states 
\begin{equation}
\mathcal{H}_{\mathsf{k}_1}^{\mathrm{chain}}
=f^{-1}(\mathsf{k}_1)\left[\frac{2}{U}-g(\mathsf{k}_1)\right],
\label{eq: effective chain Hamiltonian}
\end{equation}
provided that $f(\mathsf{k}_1)$ is invertible for all $\mathsf{k}_1\in[-\pi/\mathsf{a},\pi/\mathsf{a})$.

\begin{figure}[t]
\includegraphics[width=0.99\columnwidth]{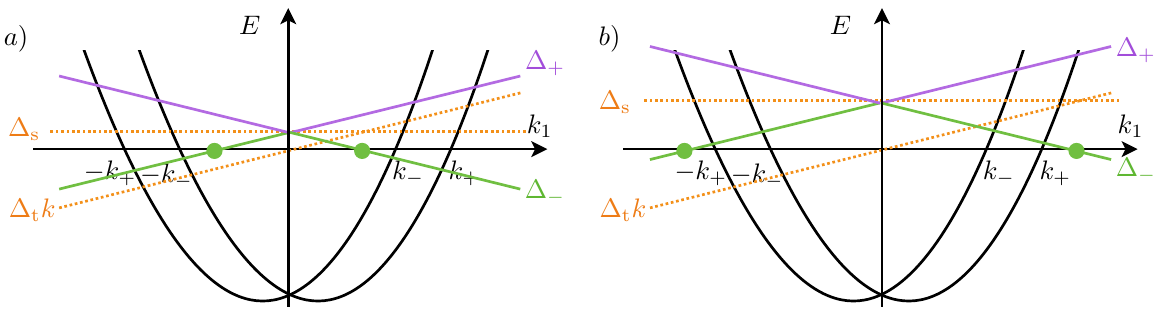}
\caption{(Color online) 
Normal state band structure of the 2D superconductor with Rashba-split bands as defined in Eq.~\eqref{eq: H0 DIII}. 
When a mixture of singlet and triplet pairing gap is induced, the superconductor can be in a) a topological phase or b) a trivial phase, depending on the relative position of the nodes of $\Delta_\lambda=\Delta_{\mathrm{s}}+\lambda|\bs{k}|\Delta_{\mathrm{t}}$ (green dots) with respect to the Fermi momentum $k_\lambda$.  For the topological properties of a scalar Shiba chain, the relative position of these nodes with respect to the Fermi momentum in the chain, rather than that of the 2D superconductor, is crucial.
        }
\label{fig: band structure 2D TRS SC}
\end{figure}

So far, we have outlined the general derivation without assuming anything about the band structure or pairing potential of the underlying superconductor. We now particularize to the continuum limit of a 2D system with Rashba spin-orbit coupling of strength $\alpha$ and a superconducting pairing potential that mixes an $s$-wave spin-singlet component $\Delta_{\mathrm{s}}$ with a $p$-wave spin-triplet component $\Delta_{\mathrm{t}}$.
The Hamiltonian
\begin{equation}
\begin{split}
\mathcal{H}^{\ }_{\boldsymbol{k}}=&\,
\left(
\frac{\bs{k}^2}{2m}
-
\mu
\right)
X^{\ }_{03}
+
\alpha\,
\left(k_2 X^{\ }_{13} - k_1 X^{\ }_{23}\right)
\\
&\,
+
\Delta^{\ }_{\text{s}}
X^{\ }_{01}
+
\Delta^{\ }_{\text{t}}
\left(k_2 X^{\ }_{11} - k_1 X^{\ }_{21}\right)
\end{split}
\label{eq: H0 DIII}
\end{equation}
has the eigenvalues
\begin{equation}
E^{2}_{\boldsymbol{k};\lambda}=
\left(
\frac{\bs{k}^2}{2m}
-
\mu
+
\lambda
\alpha|\bs{k}|
\right)^{2}
+
\left(
\Delta^{\ }_{\text{s}}
+
\lambda
\Delta^{\ }_{\text{t}}|\bs{k}|
\right)^{2},
\label{eq: 4 BdG eigenvalues}
\end{equation}
with $\lambda=\pm$. 
The TRS represented by $\mathcal{T}=X_{20}\mathcal{K}$ and the particle hole symmetry $\mathcal{P}=X_{22}\mathcal{K}$ square to $-1$ and $+1$, respectively ($\mathcal{K}$ is complex conjugation). This places the model in class DIII in the classification of Refs.~\cite{Schnyder08,Kitaev09} with a $\mathbb{Z}_2$ topological classification in 2D.
If
$\Delta^{\ }_{\text{s}}
=
-\lambda
\Delta^{\ }_{\text{t}}\,k_{\lambda}
$,
a gap-closing phase transition between the trivial and topological phase of the superconductor occurs (see Fig.~\ref{fig: band structure 2D TRS SC}). The trivial phase is the one that is adiabatically connected to the limit $|\Delta_{\mathrm{t}}|\to 0$.
Here, $k_\lambda:=\sqrt{2m\mu+m^2\alpha^2}+\lambda m \alpha$ are the Fermi momenta of the normal state band structure.
In the limit $\alpha\to0$ and $\Delta_{\mathrm{s}}\to0$, this model reduces to two copies of spinless chiral $p$-wave superconductors. Following our analysis of the full model, we will discuss this limit more specifically. We emphasize that this is a simplified, low energy effective model. The strong-coupling from of the gap can be quite different and influence our results.

In passing from the underlying 2D superconductor Hamiltonian~\eqref{eq: H0 DIII} to the effective 1D Shiba-chain Hamiltonian~\eqref{eq: effective chain Hamiltonian}, the integration over $k_2$ will eliminate all terms odd in $k_2$. These will be the terms proportional to the Pauli matrix $\sigma_1$.
To understand this, we observe that the 2D model has a mirror-symmetry $\bs{e}_2\to-\bs{e}_2$. 
For the Hamiltonian~\eqref{eq: H0 DIII}, this mirror-symmetry acting in spin and orbital space simultaneously, is represented by $\mathcal{H}^{\ }_{k_1,k_2}\to X_{20}\mathcal{H}^{\ }_{k_1,-k_2}X_{20}$. (The impurity Hamiltonian~\eqref{eq: impurity Hamiltonian} is also explicitly invariant under the mirror operation.) This representation of the mirror symmetry also holds for the inverse of $\mathcal{H}^{\ }_{\bs{k}}$ that enters the effective Hamiltonian $\mathcal{H}_{\mathsf{k}_1}^{\mathrm{chain}}$ via Eq.~\eqref{eq: R with the inverse of H}.
Since the effective Hamiltonian~\eqref{eq: effective chain Hamiltonian} is obtained by integration over $k_2$, the action of the mirror operation on it reduces to $\mathcal{H}_{\mathsf{k}_1}^{\mathrm{chain}}\to X_{20}\mathcal{H}_{\mathsf{k}_1}^{\mathrm{chain}} X_{20} $. This is equivalent to the conservation of spin in the 2-direction. The effective Hamiltonian~\eqref{eq: effective chain Hamiltonian} then only contains terms proportional to $\sigma_0$ and $\sigma_2$, restoring a $U(1)$ spin-rotation symmetry around the $\sigma_2$ axis.
It is a consequence of this extra $U(1)$ symmetry that $\mathcal{H}_{\mathsf{k}_1}^{\mathrm{chain}}$ is also invariant under the time-reversal symmetry
of spinless fermions $\mathcal{T}'=X_{20}\mathcal{T}=X_{00}\mathcal{K}$ which squares to $+1$. This extra symmetry elevates $\mathcal{H}_{\mathsf{k}_1}^{\mathrm{chain}}$ from the symmetry class DIII of the underlying system to symmetry class BDI, which features a $\mathbb{Z}$ classification in 1D. 
At the same time, it provides a simple way to determine the topological index of the model. The index of class BDI is an integer winding number $w$ that determines the number of Kramers pairs of Majorana end states. In class DIII, the $\mathbb{Z}_2$ index is the parity of this winding number. 
A similar shift of the symmetry class also appears in magnetic Shiba chains realized experimentally in Ref.~\cite{Nadj-Perge14}.   
%Thus, if we find a phase in which $w$ is odd, we can conclude that a single Kramers pair of Majorana end states will be robust against any $\sigma_2$-breaking perturbation that does not close the energy gap and preserves the physical TRS $\mathcal{T}$.

\begin{subequations}
In the eigenspaces of $\sigma_2$ with eigenvalue $s_2=\pm1$ the effective Hamiltonian reads
$\mathcal{H}_{\mathsf{k}_1;s_2}^{\mathrm{chain}}=
\bs{d}(\mathsf{k}_1,s_2)\cdot\bs{\tau}/f_{\mathsf{k}_1;s_2}$
 (a $\tau_0$-term would not respect $\mathcal{T}$ and $\mathcal{P}$),
where 
\begin{equation}
\bs{d}(\mathsf{k}_1,s_2)=\left(
\Delta_{\mathrm{s}}f_{\mathsf{k}_1;s_2}+\Delta_{\mathrm{t}}\tilde{g}_{\mathsf{k}_1;s_2},
0,
2/U-g_{\mathsf{k}_1;s_2}
\right)^{\mathsf{T}}
\label{eg: defintion d}
\end{equation}
is defined in terms of the scalar functions
\begin{equation}
\begin{split}
f_{\mathsf{k}_1;s_2}
:=&
\sum_n
\sum_{\lambda=\pm}
\int\frac{\mathrm{d}k_2}{2\pi}
\frac{1+s_2\lambda k_1/|\bs{k}|}{2E^2_{\bs{k};\lambda}},
\\
g_{\mathsf{k}_1;s_2}
:=&\sum_n\sum_{\lambda=\pm}
\int\frac{\mathrm{d}k_2}{2\pi}
\left(\frac{\bs{k}^2}{2m}-\mu+\lambda\alpha|\bs{k}|\right)
\frac{1+s_2\lambda k_1/|\bs{k}|}{2E^2_{\bs{k};\lambda}},
\\
\tilde{g}_{\mathsf{k}_1;s_2}
:=&\sum_n\sum_{\lambda=\pm}
\int\frac{\mathrm{d}k_2}{2\pi}
\frac{\lambda|\bs{k}|+s_2 k_1}{2E^2_{\bs{k};\lambda}},
\end{split}
\label{eq: f,g,tilde g}
\end{equation}
and we have implied the notation $\bs{k}=(k_1,k_2)=(\mathsf{k}_1+2\pi n/\mathsf{a},k_2)$ on the righthand sides of the equations. Notice that the factor $f_{\mathsf{k}_1;s_2}$, is manifestly positive for all $\mathsf{k}_1$.
\label{eq: hamiltonian s+p-wave chain}
\end{subequations}

To simplify matters, we would like to consider the Hamiltonian~\eqref{eq: hamiltonian s+p-wave chain} in the limit of infinitesimal superconducting gaps $\Delta_{\mathrm{s}}$ and $\Delta_{\mathrm{t}}$. To lowest order, the Hamiltonian only depends on the sign 
$\gamma_\lambda:=\mathrm{sgn}\left(\Delta_{\mathrm{s}}-\lambda\Delta_{\mathrm{t}}k_\lambda\right)$
of the pairing potential on the two Fermi surfaces at momenta $k_\lambda$, for (see~\cite{Supp} for the derivation)
\begin{subequations}
\label{eq: g and f in generality}
\begin{eqnarray}
d_1(\mathsf{k}_1,s_2)
&\approx&
m
\sum_{n}\sum_{\lambda=\pm}
\frac{\gamma_{\lambda}}{k_++k_-}\,
\frac{k_\lambda-s_2\lambda k_1}{\sqrt{k_{\lambda}^2-k_1^2}},
\label{eq: delta for infinitesimal gap}
\\
g_{\mathsf{k}_1;s_2}&\approx& 
\frac{m}{k_++k_-}
\sum_{n}\sum_{\lambda,\lambda'=\pm}
\mathrm{Re}\Biggl\{
\frac{ k_{\lambda}+s_2\lambda\,k_1}{\sqrt{k_1^2-k_{\lambda}^2}}\,
\label{eq: g for infinitesimal gap}
\\
&&
\qquad\qquad\times
\left[
\frac12
+
\frac{\lambda\lambda'}{\pi}
\mathrm{arctan}\left(
\frac{k_{\lambda}}{\sqrt{k_1^2-k_{\lambda}^2}}
\right)
\right]
\Biggr\}
%\\
%&&
%\qquad\qquad\times
%\frac{
%\frac{1}{2}+\frac{\lambda\lambda'}{\pi}\arctan\frac{k_{\lambda'}}{\sqrt{|k_{\lambda'}^2-k_1^2|}}
%}{(k_++k_{-})\sqrt{|k_{\lambda'}^2-k_1^2|}},
\nonumber
\end{eqnarray}
where it is understood that only the terms with $k_1^2<k_{\lambda}^2$ contribute to the sum for $d_1(\mathsf{k}_1,s_2)$. The $n$-dependence of the sum rests in  $k_1=\mathsf{k}_1+2\pi n/\mathsf{a}$. Formally, the sum in$g_{\mathsf{k}_1;s_2}$ is ultraviolet divergent. Introducing a Debye-frequency cutoff for the superconducting interaction energy scale much larger than $\Delta_{\mathrm{s}}$, and $\Delta_{\mathrm{t}}k_+$ and much smaller than $\mu$, is the physical cure to this divergence. The square-root singularities in Eqs.~\eqref{eq: g for infinitesimal gap} and~\eqref{eq: delta for infinitesimal gap} are regulated if higher-order contributions in $\Delta_{\mathrm{t}}$ are considered.
\end{subequations}

%\begin{figure}[t]
%\includegraphics[width=0.95\columnwidth]{Phasediagram_TRS_winding.pdf}
%\caption{(Color online) 
%Phase diagram for the 1D chain of nonmagnetic impurities on a 
%superconductor with Rashba spin-orbit coupling. The spectrum of the superconductor is described by the Rashba-spilt Fermi momenta $k_+$ and $k_-$. The color code represents the range of impurity strengths $U$ for which the chain becomes a topological superconductor with winging number $w=1$, i.e., with one Kramers pair of Majorana end modes. The superconducting pairing potential is taken to be $\Delta_{\mathrm{t}}m=0.15$, $\Delta_{\mathrm{s}}=0$. We note that for every combination $k_+$ and $k_-$, there is a range of $U$ for which the chain is topological, provided that the underlying superconductor is topological.
%Figure~\ref{fig: phase diag} presents a cut along $k_+-k_-=0$ in this phase diagram.
%        }
%\label{fig: phase diag TRS}
%\end{figure}

Let us now explore the topological properties of the 1D chain. 
As the form of Hamiltonian~\eqref{eq: hamiltonian s+p-wave chain} suggests, the winding number $w$ equals the number of windings of $\bs{d}(\mathsf{k}_1,s_2)$ around the origin of the 1-3-plane as $\mathsf{k}_1$ changes from $-\pi/\mathsf{a}$ to $\pi/\mathsf{a}$. A prerequisite for a nonvanishing winding number is that both components $d_1(\mathsf{k}_1,s_2)$ and $d_3(\mathsf{k}_1,s_2)$ change sign as a function of $\mathsf{k}_1\in[-\pi/\mathsf{a},\pi/\mathsf{a})$.
In particular, the expansions in Eq.~\eqref{eq: delta for infinitesimal gap} can be analyzed in this light.
If $\gamma_+=\gamma_-$, which is precisely the condition for the underlying 2D superconductor to be topologically trivial,
the lowest order expansion of $d_1(\mathsf{k}_1,s_2)$, Eq.~\eqref{eq: delta for infinitesimal gap}, does not change sign as function of $\mathsf{k}_1$. 
In contrast, when $\gamma_+=- \gamma_-$, i.e., if the underlying 2D superconductor is topologically nontrivial, the summand in Eq.~\eqref{eq: delta for infinitesimal gap} diverges to $(\gamma_{\lambda}\infty)$, when $k_1\to -\lambda s_2 k_\lambda$. 
If we include the folding of the momentum to the Brillouin zone $\mathsf{k}_1\in[-\pi/\mathsf{a},\pi/\mathsf{a})$, this means
 that $d_1(\mathsf{k}_1,s_2)$ diverges to $(\gamma_+\infty)$ at $\mathsf{k}_{1,+}:=(-s_2k_+)\mathrm{mod}\, 2\pi/\mathsf{a}$ and to $(\gamma_-\infty)$ at $\mathsf{k}_{1,-}:=(s_2k_-)\mathrm{mod} \,2\pi/\mathsf{a}$ . Thus, $d_1(\mathsf{k}_1,+)$ has to change sign between $\mathsf{k}_{1,-}$ and $\mathsf{k}_{1,+}$ if $\gamma_+=- \gamma_-$.  Due to the $2\pi/\mathsf{a}$ periodicity in $\mathsf{k}_1$, $d_1(\mathsf{k}_1,s_2)$ has thus at least two zeros in the interval $\mathsf{k}_1\in[-\pi/\mathsf{a},\pi/\mathsf{a})$. Denote these zeros by $\mathsf{k}_{1,0}$ and $\mathsf{k}^\prime_{1,0}$, and let us consider the case where these are the only zeros of $d_1(\mathsf{k}_1,s_2)$. For the winding number $w$ to be nontrivial, $d_3(\mathsf{k}_1,s_2)$ has to change sign between these two zeros. Generically, the function $g_{\mathsf{k}_1,s_2}$ that enters $d_3(\mathsf{k}_1,s_2)$ will take different values at the two zeros. Let us assume, without loss of generality, $g_{\mathsf{k}_{1,0},s_2}< g_{\mathsf{k}^\prime_{1,0},s_2}$. 
 Then, according to Eq.~\eqref{eg: defintion d}, if 
 $U\in\left(2/g_{\mathsf{k}^\prime_{1,0},s_2},2/g_{\mathsf{k}_{1,0},s_2}\right)$
the component $d_3(\mathsf{k}_1,s_2)$ changes sign between the two zeros and the winding number is one. 
Our considerations thus show how the impurity chain becomes a nontrivial 1D superconductor,
if the underlying superconductor is topological, $U$ lies in the appropriate range and $d_1(\mathsf{k}_1,s_2)$ has only two zeros. If one of these conditions is not met, however, the chain may remain in a topologically trivial state. 

A natural question that follows from these observations is whether the nontrivial topology of the underlying superconductor is a \emph{prerequisite} for the impurity chain to become topological. We will now argue that this is not the case. First, we show analytically that Eqs.~\eqref{eq: f,g,tilde g} allow for this scenario in a certain limit. Second, we numerically solve finite systems and show the existence of Majorana bound states in a chain of scalar impurities on a trivial superconductor. 

For the analytical argument, we consider the simplified case $k_+=k_-$ and the limit $\mathsf{a} k_\pm\ll\pi$. 
By treating the underlying 2D superconductor in the continuum limit, we have assumed that its lattice constant $a$ is much smaller than that of the chain $\mathsf{a}\gg a$. In a lattice description of the 2D superconductor, its Brillouin zone $[-\pi/a,\pi/a)$ is thus much larger than that of the chain  $[-\pi/\mathsf{a},\pi/\mathsf{a})$. The constraint that the Fermi momenta lie in the smaller chain Brillouin zone $\mathsf{a} k_\pm\ll\pi$ is thus satisfied if the chemical potential of the superconductor is sufficiently low.
Then, the expansion of $d_1(\mathsf{k}_1,s_2)$ to lowest (zeroth) order in the gap functions, as given by Eq.~\eqref{eq: delta for infinitesimal gap}, vanishes identically for $k_\pm<|\mathsf{k}_1|<\pi/\mathsf{a}$ (while it has definite sign $\gamma_+=\gamma_-$ otherwise).
We thus have to appeal to the next (linear) order corrections in the superconducting gaps to this result, 
and in particular ask whether they allow for a sign change of $d_1(\mathsf{k}_1,s_2)$ for $\mathsf{k}_1$ with $k_\pm<|\mathsf{k}_1|<\pi/\mathsf{a}$. 
In the limit $\mathsf{a} k_\pm\ll \pi$, due to the denominator, it is sufficient to restrict the sums over $n$ 
 in Eqs.~\eqref{eq: f,g,tilde g} to the $n=0$ term. 
Further, if we also restrict us to small $|\mathsf{k}_1|\ll \pi/\mathsf{a}$, we can expand the terms in $\mathsf{k}_1$, with the understanding that $ k_\pm< \mathsf{k}_1\ll \pi/\mathsf{a}$, and obtain
%\begin{equation}
$d_1(\mathsf{k}_1,s_2)\approx m^2(\Delta_{\mathrm{s}}+s_2\mathsf{k}_1\Delta_{\mathrm{t}})/|\mathsf{k}_1|^3$,
%\end{equation}
as well as $g_{\mathsf{k}_1;s_2}\approx |\mathsf{k}_1|^{-1}$. 
We observe that $d_1(\mathsf{k}_1,s_2)$ indeed changes sign at momentum $\mathsf{k}_1^{(0)}= -s_2\Delta_{\mathrm{s}}/\Delta_{\mathrm{t}}$. (Note that the condition to have a trivial underlying superconductor $\gamma_+=\gamma_-$ implies $|\mathsf{k}_1^{(0)}|>k_\pm$.)
One can take advantage of this sign change in $d_1$ by choosing $U$ in such a way that the vector $\bs{d}(\mathsf{k}_1,s_2)$ winds. Thus, the chain of scalar impurities can be topologically nontrivial, hosting a Kramers pair of Majorana bound states at its end, even if the underlying superconductor is topologically trivial. 

This mathematical result has an intuitive understanding. 
If the 2D superconductor has a sizeable triplet pair potential, 
there is a circle at $|\bs{k}|=|\Delta_{\mathrm{s}}/\Delta_{\mathrm{t}}|$ in the $k_1$-$k_2$-plane along which the gap function is nodal. If the 2D superconductor is topologically trivial, the Fermi surfaces of the metallic state before pairing lie inside this nodal circle, i.e., the Fermi momenta satisfy $k_\pm<|\Delta_{\mathrm{s}}/\Delta_{\mathrm{t}}|$. The impurity chain, on the other hand, will have its own independent Fermi momentum $k_{\mathrm{F,1D}}$. The chain is in a nontrivial state if the paring potential is of opposite sign on the two Fermi points of the chain, that is, if the nodal line of the order parameter lies between the 2D and 1D Fermi momenta 
$k_\pm<|\Delta_{\mathrm{s}}/\Delta_{\mathrm{t}}|<k_{\mathrm{F,1D}}$. We note that this is a one-body consideration beyond the weak pairing limit (requires the superconducting gaps to be well formed for a larger $k$-space region than just the superconductor Fermi surfaces), and might not survive a self-consistent calculation.

\begin{figure}[t]
\includegraphics[width=0.99\columnwidth]{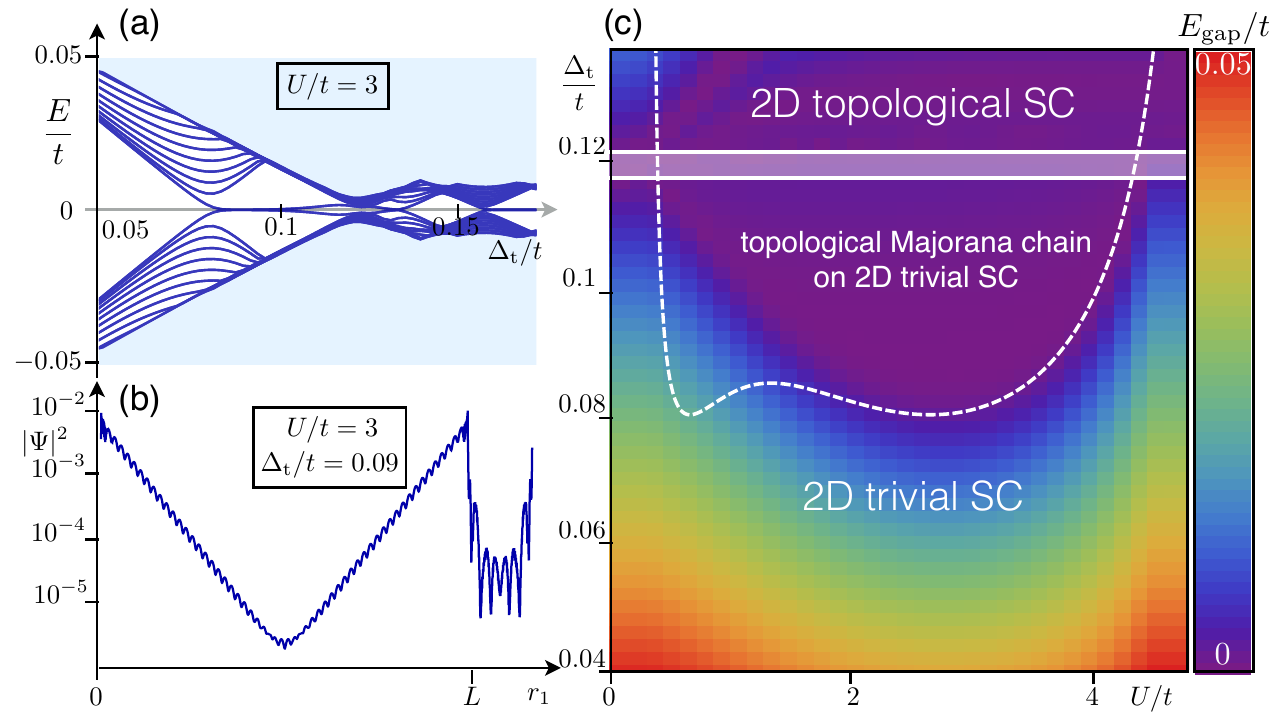}
\caption{
Topological chain of scalar impurities on a non-topological 2D superconductor.
(a) Lowest eigenvalues of Hamiltonian~\eqref{eq: lattice Hamiltonian} for 
$\mu/t=3.5$, 
$\Delta_{\mathrm{s}}/t=0.16$,  
$L_1\times L_2=2000\times 26$, $L=1800$. Between $\Delta_{\mathrm{t}}/t\sim 0.08$ and $\Delta_{\mathrm{t}}/t\sim 0.12$ the chain is in a topological and the underlying superconductor are in a trivial phase. (b) The Kramers pair of Majorana modes at each end of the chain is exponentially localized
(c) Phase diagram as a function of the triplet pairing potential $\Delta_{\mathrm{t}}$ and the strength of the scalar impurities $U$, with the color scale representing the lowest energy eigenvalue of Hamiltonian~\eqref{eq: lattice Hamiltonian}. Thus, blue regions in the phase diagram either indicate the presence of Majorana end states or a gapless bulk superconductor. The shaded region between the solid lines marks the topological phase transition of the 2D bulk superconductor. The dashed line shows an analytical estimate for the topological phase transition in the chain. It is obtained by restricting Hamiltonian~\eqref{eq: lattice Hamiltonian} to the lattice sites one the chain only, with the inclusion of second order virtual processes that correspond to hopping one site away from the chain and back. 
        }
\label{fig: nontrivial on trivial}
\end{figure}

We have obtained the analytical result as a proof of principle in the limit $|\mathsf{k}_1|\ll \pi/\mathsf{a}$ in order to make the calculation tractable. However, the general statement that nontrivial chain topology can emerge out of a trivial substrate (in BdG formalism) is not restricted to this limit, as we now show numerically. 
To that end, consider the BdG Hamiltonian 
\begin{equation}
\begin{split}
H
=
&
\sum_{\bs{r}\in\Lambda}\sum_{s=\pm}\sum_{j=1,2}
\left[
t\,c^\dagger_{\bs{r},s}c^{\ }_{\bs{r}+\bs{e}_j a,s}
+
(\mathrm{i}s)^j\Delta_{\mathrm{t}}\,c^\dagger_{\bs{r},s}c^{\dagger}_{\bs{r}+\bs{e}_ja,s}
\right]
\\
&
+\sum_{\bs{r}\in\Lambda}
\left[\sum_{s=\pm}
\frac{U_{\bs{r}}-\mu}{2}\,c^\dagger_{\bs{r},s}c^{\ }_{\bs{r},s}
+
\Delta_{\mathrm{s}}\,c^\dagger_{\bs{r},\uparrow}c^{\dagger}_{\bs{r},\downarrow}
\right]
+\mathrm{h.c.}.
\end{split}
\label{eq: lattice Hamiltonian}
\end{equation} 
defined on a square lattice $\Lambda=L_1\times L_2$, spanned by the vectors $\bs{e}_1a$ and  $\bs{e}_2a$, with periodic boundary conditions
Here, $c^\dagger_{\bs{r},s}$ creates an electron of spin $s$ at site $\bs{r}$, while $t$ and $\mu$ are the nearest neighbor hopping integrals and the chemical potential, respectively. 
The local potential $U_{\bs{r}}$ takes the nonzero value $U$ along the Shiba chain of length $L<L_1$, say for $r_1/a=0,\cdots, L$ and $r_2=0$, while it vanishes otherwise.
Figure~\ref{fig: nontrivial on trivial} shows that for a choice of $t$ and $\mu$, for which the 2D Fermi surface is small, there exists a Kramers pair of exponentially localized Majorana bound states at each end of the chain for values of the pairing potential $\Delta_{\mathrm{t}}$ that lie below the bulk trivial to  topological superconductor phase transition.

\begin{figure}[t]
\includegraphics[width=0.9\columnwidth]{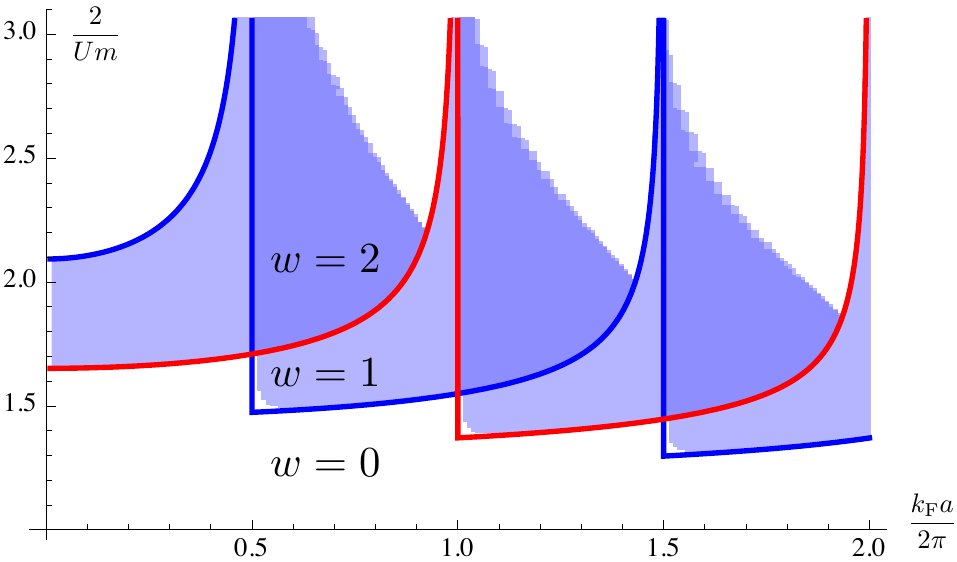}
\caption{
Phase diagram for the 1D chain of nonmagnetic impurities on a chiral $p$-wave superconductor as a function of the strength of the impurity potential $U$ and the spacing $\mathsf{a}$ of the impurity atoms in units of the Fermi wavelength $2\pi/k_{\mathrm{F}}$ of the chiral superconductor.
The colored regions correspond to different values of the topological index $w$ of the 1D chain, which has been evaluated numerically in presence of a small regulating gap $\Delta_{\mathrm{t}}$. Physically, $w$ equals the number of Majorana end modes of the chain. 
        }
\label{fig: phase diag}
\end{figure}

Before closing, let us discuss topological properties of a chain of scalar impurities on a
chiral $p$-wave superconductor of spineless fermions. The Hamiltonian for this system
is readily obtained by evaluating Hamiltonian~\eqref{eq: hamiltonian s+p-wave chain} for one spin species (say $s_2=+1$) in the limit $\alpha\to0$ and $\Delta_{\mathrm{s}}\to0$.
In this limit, one obtains a series of transitions between phases $w=1$ and with even $w$, maked by the phase transition lines in see Fig.~\ref{fig: phase diag} as a function of the impurity spacing $\mathsf{a}$.~\cite{Supp}
 An explicit numerical evaluation of the winding number $w$ reveals an even richer structure including regions in the $U$--$(k_{\mathrm{F}}\mathsf{a})$ phase diagram in which $w=2$. We explicitly checked that the Hamiltonian is gapped for all the phases presented in Fig.~\ref{fig: phase diag}.

In summary, we have shown that a chain of scalar impurities that is deposited on fully gapped unconventional superconductors can be a topologically nontrivial 1D superconductor. We have exemplified this for the case of a chiral $p$-wave superconductor and for a Rashba spin-orbit coupled superconductor with both singlet and triplet pairing potential. In the latter case, the impurity chain can be in a nontrivial phase, even if the underlying superconductor is topologically trivial. However, for this to happen, the oder parameter must be nodal for momenta outside the Fermi sea -- not necessarily a generic situation. 

We hope our results will stimulate work on nonmagnetic impurities on Sr$_2$RuO$_4$.
The system we considered may naturally occur at step edges~\cite{Misra02,Zhoun13}
 and related linear defects on the surface of such chiral superconductors. Our results suggest that the ends of such linear defects could carry interesting bound states.

\begin{acknowledgments}
This work was supported by NSF CAREER DMR-095242, ONR - N00014-11-1-0635, ARO MURI W911NF-12-1-0461, NSF-MRSEC DMR-0819860, Packard Foundation and Keck grant. 
  \end{acknowledgments}

\clearpage
\newpage
%\bibliography{bib}

\begin{thebibliography}{99}
\bibitem{footnote-3D}
The same derivation would hold for a 3D superconductor, with quantitatively different results.
\end{thebibliography}

\appendix
\onecolumngrid
\vspace{\columnsep}

\section{SUPPLEMENTAL MATERIAL}

\section{Parity transitions in impurity states}
In this section, we would like to give an intuitive picture as to why impurity sites in $s$ and $p$-wave superconductors can bind an impurity states (the Shiba states) and why those sites can experience a topological transition, at which the parity of the many-body ground state changes. To this end, we consider the simplest noninteracting model of a single site and a pair of sites, respectively, but work in its many-body Hilbert space.

In symmetry class $D$ (applies to chiral $p$-wave superconductors) the $\mathbb{Z}_2$ topological index of a 1D superconductor denotes the change in fermion parity of the ground-state as a flux $\pi$ is inserted through a system with periodic boundary conditions. This parity change can even be observed in zero-dimensional models of isolated impurities and provides basic intuition whether and how a 1D chain of scalar impurities has the potential to undergo a topological phase transition. 
Here, we discuss this minimal model for a parity changing transition for one and two sites populated with spinless fermions and for a single site populated with spinful fermions. These trivial models provide insight into why in-gap Shiba states can be present in scalar impurities placed in superconductors, complementing the more well known situation of Shiba states occurring in magnetic impurities.

\subsubsection{Spinless fermions}
A single spinless fermion cannot exhibit superconducting pairing. Irrespective of that, we see that an on-site chemical potential $\mu$ can change the parity $P$ of the ground state, for the latter is given by $\mathrm{sgn} \mu$. In the occupation basis $(|0\rangle,|1\rangle)$, the Hamiltonian reads 
\begin{sequation}
H=
\begin{pmatrix}
0&0\\
0&-\mu
\end{pmatrix}.
\label{eq: Hamham}
\end{sequation}
Any other terms in the Hamiltonian violate the conservation of Fermion parity.
The ground state is given by
$|0\rangle$ and $|1\rangle$ for $\mu<0$ and $\mu>0$, respectively, with opposite parity.

The minimal extension to this model that includes superconducting pairing includes two sites with spinless fermions. In this case, we can have triplet -- but not singlet -- superconducting pairing. 
In the basis $(|0,0\rangle,|1,0\rangle,|0,1\rangle,|1,1\rangle)$, the Hamiltonian reads 
\begin{sequation}
H=
\begin{pmatrix}
0&0&0&\Delta\\
0&-\mu&t&0\\
0&t&-\mu&0\\
\Delta&0&0&-2\mu
\end{pmatrix},
\end{sequation}
where $t$ is the hopping integral between the two sites.
The energies are
\begin{sequation}
\varepsilon^{\text{even}}=\pm\sqrt{\Delta^2+\mu^2}-\mu,
\qquad
\varepsilon^{\text{odd}}=\pm t-\mu.
\end{sequation}
Whenever $|t|>\Delta$ (commonly referred to as the weak pairing phase), we can induce a parity change of the ground state (protected crossing) by changing the chemical potential at
\begin{sequation}
\mu^2=t^2-\Delta^2.
\end{sequation} 
For smaller $|\mu|$, the ground state has odd parity, for larger $|\mu|$, it has even parity.
This is in line with the behavior of bound states of two impurities in $p$-wave superconductors: They exhibit a protected crossing in the bound state spectrum.~\cite{Choi04}. The presence of this protected crossing implies the existence of sub-gap Shiba states in the energy spectrum: since a protected crossing has to occur upon varying $\mu$, which could be considered as modeling the scalar impurity strength, it must be that sub-gap $E<\Delta$ states exist - these are the Shiba states. 

\subsubsection{Spinful fermions}
A single site with a spinful fermion degree of freedom we exhibit singlet superconducting pairing $\Delta$. We also apply a Zeeman field $B$ in the direction of the spin-quantization axis. 
In the basis $(|0,0\rangle,|\uparrow,0\rangle,|0,\downarrow\rangle,|\uparrow,\downarrow\rangle)$, the Hamiltonian reads 
\begin{sequation}
H=
\begin{pmatrix}
0&0&0&\Delta\\
0&-\mu+B&0&0\\
0&0&-\mu-B&0\\
\Delta&0&0&-2\mu\\
\end{pmatrix}.
\end{sequation}
The energies are
\begin{sequation}
\varepsilon^{\text{even}}=\pm\sqrt{\Delta^2+\mu^2}-\mu,
\qquad
\varepsilon^{\text{odd}}=\pm B-\mu.
\end{sequation}
We observe a level crossing protected by parity symmetry at
\begin{sequation}
B^2=\Delta^2+\mu^2.
\end{sequation}
For smaller $|B|$, the ground state has even parity, for larger $|B|$, it has odd parity.
This is congruent with the behavior of a ferromagnetic Shiba chain on an $s$-wave superconductor. 
This model also shows that a density impurity cannot induce a subgap bound state deep in an $s$-wave superconducting gap, for $\mu$ does not induce any phase transition in this model for $B=0$.

\section{Derivation of Eq.~\eqref{eq: g for infinitesimal gap} and Eq.~\eqref{eq: delta for infinitesimal gap}}

Starting from Eq.~\eqref{eq: f,g,tilde g}, we want to consider the limit of small $\Delta_{\mathrm{s}}$ and $\Delta_{\mathrm{t}}$.
We rewrite
\begin{sequation}
E^{2}_{\boldsymbol{k};\lambda}=\frac{1}{(2m)^2}
\left[
\left(|\bs{k}|+k_\lambda\right)^2
\left(|\bs{k}|-k_{-\lambda}\right)^2
+
\delta_{\lambda,|\bs{k}|}^{2}\right],
\end{sequation}
where
\begin{sequation}
\delta_{\lambda,|\bs{k}|}
:=2m(\Delta^{\ }_{\text{s}}+\lambda\Delta^{\ }_{\text{t}}|\bs{k}|).
\end{sequation}
We will need the following two integrals. We are only interested in the lowest order nonvanishing terms in $\delta_{\lambda,|\bs{k}|}$, which is order $\mathcal{O}(1/\delta_{\lambda,|\bs{k}|})$ for the first integral and $\mathcal{O}(1)$ for the second integral. 
\begin{sequation}
\begin{split}
I_{1,-\lambda}(k_1):
=&\int\frac{\mathrm{d}k_2}{2\pi}\frac{1}{(\sqrt{k_1^2+k_2^2}-k_\lambda)^2(\sqrt{k_1^2+k_2^2}+k_{-\lambda})^2+\delta_{-\lambda,k_\lambda}^2}
\\
=&
\frac{1}{2\mathrm{i}\delta_{-\lambda,k_\lambda}}
\sum_{\gamma=\pm}
\gamma
\int\frac{\mathrm{d}k_2}{2\pi}
\frac{1}{
(\sqrt{k_1^2+k_2^2}-k_\lambda)(\sqrt{k_1^2+k_2^2}+k_{-\lambda})-\mathrm{i}\gamma\delta_{-\lambda,k_\lambda}
}
\\
=&
\frac{1}{2\mathrm{i}\delta_{-\lambda,k_\lambda}}
\sum_{\gamma=\pm}
\gamma
\int\frac{\mathrm{d}k_2}{2\pi}
\frac{1}{
\left(\sqrt{k_1^2+k_2^2}-k_\lambda-\mathrm{i}\frac{\gamma\delta_{-\lambda,k_\lambda}}{k_++k_-}\right)\left(\sqrt{k_1^2+k_2^2}+k_{-\lambda}+\mathrm{i}\frac{\gamma\delta_{-\lambda,k_\lambda}}{k_++k_-}\right)
+\mathcal{O}(\delta_{-\lambda,k_\lambda}^2)
}
\end{split}
\label{eq: I1}
\end{sequation}
We compute the integral (for $\kappa_\lambda:=k_\lambda/|k_1|>1$)
\begin{sequation}
\begin{split}
&\int\mathrm{d}\kappa\frac{1}{
\left(\sqrt{1+\kappa^2}-\kappa_\lambda-\mathrm{i}\Delta\right)
\left(\sqrt{1+\kappa^2}+\kappa_{-\lambda}+\mathrm{i}\Delta\right)}
\\
&\qquad=\frac{1}{\kappa_++\kappa_-+2\mathrm{i}\Delta}
\sum_{\lambda'=\pm}
\frac{(\kappa_{\lambda'\lambda}+\mathrm{i}\Delta)}{\sqrt{1-(\kappa_{\lambda'\lambda}+\mathrm{i}\Delta)^2}}
\left[
\arctan\left(\frac{\kappa}{\sqrt{1-(\kappa_{\lambda'\lambda}+\mathrm{i}\Delta)^2}}\right)\right.
\\
&\qquad\qquad\left.
+\lambda'
\arctan\left(\frac{\kappa(\kappa_{\lambda'\lambda}+\mathrm{i}\Delta)}{\sqrt{1+\kappa^2}\sqrt{1-(\kappa_{\lambda'\lambda}+\mathrm{i}\Delta)^2}}\right)
\right]_{\kappa=-\infty}^{\kappa=+\infty}
\\
&\qquad
=\frac{1}{\kappa_++\kappa_-+2\mathrm{i}\Delta}
\sum_{\lambda'=\pm}
\frac{(\kappa_{\lambda'\lambda}+\mathrm{i}\Delta)}{\sqrt{1-(\kappa_{\lambda'\lambda}+\mathrm{i}\Delta)^2}}
\left[\pi (1+\lambda')
+2\lambda'\mathrm{i}\,\mathrm{sgn}(\Delta)
\mathrm{artanh}
\left(\frac{\sqrt{\kappa_{\lambda'\lambda}^2-1}}{\kappa_{\lambda'\lambda}}\right)
\right]+\mathcal{O}(\Delta)
\\
&\qquad
=\frac{\mathrm{i}\,\mathrm{sgn}(\Delta)}{\kappa_++\kappa_-}
\sum_{\lambda'=\pm}
\frac{\kappa_{\lambda'\lambda}}{\sqrt{\kappa_{\lambda'\lambda}^2-1}}
\left[\pi (1+\lambda')
+2\lambda'\mathrm{i}\,\mathrm{sgn}(\Delta)
\mathrm{artanh}
\left(\frac{\sqrt{\kappa_{\lambda'\lambda}^2-1}}{\kappa_{\lambda'\lambda}}\right)
\right]+\mathcal{O}(\Delta),
\end{split}
\label{eq: nr integral 1}
\end{sequation}
with $\mathrm{artanh}$ the inverse of the hyperbolic tangent.
Notice that $\Delta$, that will be substituted by $\gamma\delta_{-\lambda,k_\lambda}/[|k_1|(k_++k_-)]$ with $\gamma=\pm,\ \lambda=\pm$, is a dimensionless number.
Here we used
\begin{sequation}
\arctan\left(\frac{a+\mathrm{i}b}{\sqrt{1-(a+\mathrm{i} b)^2}}\right)
=\frac{\pi}{2}+\mathrm{i}\,\mathrm{sgn}(b)\,\mathrm{artanh}\left(\frac{\sqrt{a^2-1}}{a}\right)+\mathcal{O}(b)
\end{sequation}
and
\begin{sequation}
\frac{a+\mathrm{i}b}{\sqrt{1-(a+\mathrm{i} b)^2}}
=\mathrm{i}\,\mathrm{sgn}(b)\frac{a}{\sqrt{a^2-1}}+\mathcal{O}(b).
\end{sequation}
We now use Eq.~\eqref{eq: nr integral 1} to determine Eq.~\eqref{eq: I1} with 
$\kappa=k_2/|k_1|$, 
$\kappa_{\lambda'\lambda}=k_{\lambda'\lambda}/|k_1|$, and
$\Delta=\gamma\delta_{-\lambda,k_\lambda}/[|k_1|(k_++k_-)]$
for the case $k_\lambda>|k_1|$
\begin{sequation}
\begin{split}
I_{1,-\lambda}(k_1)
=&
\frac{1}{|\delta_{-\lambda,k_\lambda}|}
\frac{k_\lambda}{k_++k_{-}}
\frac{1}{\sqrt{k_\lambda^2-k_1^2}}
+\mathcal{O}\left(1\right),
\end{split}
\end{sequation}
where $\mathcal{O}(1)=\mathcal{O}\left(\left|\delta_{-\lambda,k_\lambda}/[k_1(k_++k_-)]\right|^0\right)$.
In contrast, if $k_\lambda<|k_1|$ the integral in Eq.~\eqref{eq: I1} is completely regular in the limit $\delta_{-\lambda,k_\lambda}\to 0$, so that we conclude that the leading order in $\delta_{-\lambda,k_\lambda}$ is given by
\begin{sequation}
\begin{split}
I_{1,-\lambda}(k_1)
=&
\begin{cases}
\frac{1}{|\delta_{-\lambda,k_\lambda}|}
\frac{k_\lambda}{k_++k_{-}}
\frac{1}{\sqrt{k_\lambda^2-k_1^2}}
+\mathcal{O}(1)\quad,&|k_1|<k_\lambda,\\
\mathcal{O}(1),&|k_1|>k_\lambda.
\end{cases}
\end{split}
\end{sequation}

The second integral that we need is
\begin{sequation}
\begin{split}
I_{2,-\lambda,s}(k_1):=
&
\int\frac{\mathrm{d}k_2}{2\pi}
\left(
1+s \lambda\frac{k_1}{\sqrt{k_1^2+k_2^2}}
\right)
\frac{(\sqrt{k_1^2+k_2^2}-k_\lambda)(\sqrt{k_1^2+k_2^2}+k_{-\lambda})}{(\sqrt{k_1^2+k_2^2}-k_\lambda)^2(\sqrt{k_1^2+k_2^2}+k_{-\lambda})^2+\delta_{-\lambda}^2}.
\end{split}
\end{sequation}
We can compute the leading contributions $\mathcal{O}(1)$ of this integral simply by setting $\delta_{-\lambda}=0$ yielding
\begin{sequation}
\begin{split}
I_{2,-\lambda,s}(k_1)
=&
\int\frac{\mathrm{d}k_2}{2\pi}
\left(
1+s \lambda\frac{k_1}{\sqrt{k_1^2+k_2^2}}
\right)
\frac{1}{(\sqrt{k_1^2+k_2^2}-k_\lambda)(\sqrt{k_1^2+k_2^2}+k_{-\lambda})}
+
\mathcal{O}(\delta_{-\lambda}/k_1^2)
\\
=&
\frac{1}{k_++k_-}
\sum_{\lambda'=\pm}
\mathrm{Re}\left\{
\frac{ k_{\lambda'\lambda}+s\lambda\lambda'\,k_1}{\sqrt{k_1^2-k_{\lambda'\lambda}^2}}\,
\left[
\frac12
+
\frac{\lambda'}{\pi}
\mathrm{arctan}\left(
\frac{k_{\lambda'\lambda}}{\sqrt{k_1^2-k_{\lambda'\lambda}^2}}
\right)
\right]
\right\}
+\mathcal{O}(\delta_{-\lambda}/k_1^2).
\end{split}
\end{sequation}

We will need two more equalities. They can be solved by manipulations similar to Eq.~\eqref{eq: I1}, namely
\begin{sequation}
\begin{split}
&
\int\frac{\mathrm{d}k_2}{2\pi}
\frac{1}{\sqrt{k_1^2+k_2^2}}\,\frac{1}
{(\sqrt{k_1^2+k_2^2}-k_\lambda)^2(\sqrt{k_1^2+k_2^2}+k_{-\lambda})^2+\delta_{-\lambda,k_\lambda}^2}
\\
&\qquad
=
\frac{1}{k_\lambda} I_{1,-\lambda}(k_1)
+
\frac{1}{2\mathrm{i}\delta_{-\lambda,k_\lambda}}
\sum_{\gamma=\pm}
\gamma
\int\frac{\mathrm{d}k_2}{2\pi}
\frac{k_\lambda-\sqrt{k_1^2+k_2^2}}{\sqrt{k_1^2+k_2^2}k_\lambda}
\frac{1}{
(\sqrt{k_1^2+k_2^2}-k_\lambda)(\sqrt{k_1^2+k_2^2}+k_{-\lambda})
}
+\mathcal{O}(1)
\\
&\qquad
=
\frac{1}{k_\lambda} I_{1,-\lambda}(k_1) +\mathcal{O}(1)
\end{split}
\label{eq: I1 variant 1}
\end{sequation}
and
\begin{sequation}
\begin{split}
&
\int\frac{\mathrm{d}k_2}{2\pi}
\sqrt{k_1^2+k_2^2}\,\frac{1}
{(\sqrt{k_1^2+k_2^2}-k_\lambda)^2(\sqrt{k_1^2+k_2^2}+k_{-\lambda})^2+\delta_{-\lambda,k_\lambda}^2}
\\
&\qquad
=
k_\lambda\, I_{1,-\lambda}(k_1)
+
\frac{1}{2\mathrm{i}\delta_{-\lambda,k_\lambda}}
\sum_{\gamma=\pm}
\gamma
\int\frac{\mathrm{d}k_2}{2\pi}
\frac{\sqrt{k_1^2+k_2^2}-k_\lambda}{
(\sqrt{k_1^2+k_2^2}-k_\lambda)(\sqrt{k_1^2+k_2^2}+k_{-\lambda})
}
+\mathcal{O}(1)
\\
&\qquad
=
\frac{1}{k_\lambda} I_{1,-\lambda}(k_1) +\mathcal{O}(1).
\end{split}
\label{eq: I1 variant 2}
\end{sequation}

We can now approximate the functions entering the chain Hamiltonian using Eqs.~\eqref{eq: I1 variant 1} and~\eqref{eq: I1 variant 1}
\begin{sequation}
\begin{split}
f_{\mathsf{k}_1;s_2}
&=
2m^2\sum_n
\sum_{\lambda=\pm}
\int\frac{\mathrm{d}k_2}{2\pi}
\frac{1-s_2\lambda k_1/|\bs{k}|}{
\left(|\bs{k}|-k_\lambda\right)^2
\left(|\bs{k}|+k_{-\lambda}\right)^2
+
\delta_{\lambda,|\bs{k}|}^{2}
}
\\
&=
2m^2\sum_n
\sum_{\lambda=\pm}
\int\frac{\mathrm{d}k_2}{2\pi}
\frac{1-s_2\lambda k_1/k_\lambda}{
\left(|\bs{k}|-k_\lambda\right)^2
\left(k_\lambda+k_{-\lambda}\right)^2
+
\delta_{\lambda,|\bs{k}|}^{2}
}
+\mathcal{O}(1)
\\
&=
2m^2\sum_n
\sum_{\lambda=\pm}
\left(1-s_2\lambda \frac{k_1}{k_\lambda}\right)
I_{1,\lambda}(k_1)
+\mathcal{O}(1)
\end{split}
\end{sequation}
and 
\begin{sequation}
\begin{split}
\tilde{g}_{\mathsf{k}_1;s_2}
&=
2m^2\sum_n
\sum_{\lambda=\pm}
\int\frac{\mathrm{d}k_2}{2\pi}
\frac{-\lambda|\bs{k}|+s_2 k_1}{
\left(|\bs{k}|-k_\lambda\right)^2
\left(|\bs{k}|+k_{-\lambda}\right)^2
+
\delta_{-\lambda,|\bs{k}|}^{2}
}
\\
&=
2m^2\sum_n
\sum_{\lambda=\pm}
\int\frac{\mathrm{d}k_2}{2\pi}
\frac{-\lambda k_\lambda+s_2 k_1}{
\left(|\bs{k}|-k_\lambda\right)^2
\left(k_\lambda+k_{-\lambda}\right)^2
+
\delta_{-\lambda,|\bs{k}|}^{2}
}
+\mathcal{O}(1)
\\
&=
2m^2\sum_n
\sum_{\lambda=\pm}
\left(-\lambda k_\lambda+s_2 k_1\right)
I_{1,\lambda}(k_1)
+\mathcal{O}(1)
\end{split}
\end{sequation}
as well as
\begin{sequation}
\begin{split}
g_{\mathsf{k}_1;s_2}
=&m
\sum_n\sum_{\lambda=\pm}
\int\frac{\mathrm{d}k_2}{2\pi}
\left(|\bs{k}|+k_\lambda\right)
\left(|\bs{k}|-k_{-\lambda}\right)
\frac{1+s_2\lambda k_1/|\bs{k}|}
{\left(|\bs{k}|+k_\lambda\right)^2
\left(|\bs{k}|-k_{-\lambda}\right)^2
+
\delta_{\lambda,|\bs{k}|}^{2}}
\\
=&m
\sum_n\sum_{\lambda=\pm}
\int\frac{\mathrm{d}k_2}{2\pi}
\frac{1+s_2\lambda k_1/|\bs{k}|}
{\left(|\bs{k}|+k_\lambda\right)
\left(|\bs{k}|-k_{-\lambda}\right)
}
+\mathcal{O}\left(\frac{\delta_{\lambda,|\bs{k}|}}{\left(|\bs{k}|+k_\lambda\right)
\left(|\bs{k}|-k_{-\lambda}\right)}\right)
\\
\approx&m\sum_n\sum_{\lambda=\pm}
I_{2,\lambda,-s_2}(k_1)
.
\end{split}
\label{eq: f,g,tilde g app}
\end{sequation}
In particular, the linear combination
\begin{sequation}
\begin{split}
\Delta_{\mathrm{s}}f_{\mathsf{k}_1;s_2}
+
\Delta_{\mathrm{t}}\tilde{g}_{\mathsf{k}_1;s_2}
&\approx
2m^2
\sum_{n}\sum_{\lambda=\pm}
I_{1,-\lambda}(k_1)
\left[\Delta_{\mathrm{s}}
\left(1-s_2\lambda \frac{k_1}{k_\lambda}\right)
+
\Delta_{\mathrm{t}}
\left(-\lambda k_\lambda+s_2 k_1\right)
\right]
\\
&=
m
\sum_{n}\sum_{\lambda=\pm}
I_{1,-\lambda}(k_1)
\left(1-s_2\lambda \frac{k_1}{k_\lambda}\right)
\delta_{-\lambda,k_\lambda}
\\
&\approx
m
\sum_{n}\sum_{\lambda=\pm}
\frac{k_\lambda-s_2\lambda k_1}
{k_++k_{-}}
\frac{\mathrm{sgn}\delta_{-\lambda,k_\lambda}
}{\sqrt{k_\lambda^2-k_1^2}}
\end{split}
\end{sequation}
where it is understood that only terms with real square root contribute.

\section{Nontrivial chains from trivial superconductors}

Here we provide an analytic argument that in the limit where the lattice spacing $\mathsf{a}$ of the impurity chain is much smaller than the inverse Fermi momenta $k_\pm^{-1}$ of the underlying superconductor it is possible to construct a nontrivial 1D superconducting chain on a trivial 2D superconductor. To obtain a nontrivial chain in this limit, we further require that the triplet gap $\Delta_{\mathrm{t}}$ is nonvanishing and that the strength $U$ of the scalar impurities can be appropriately tuned. 
In the limit $\mathsf{a}\ll k_\pm^{-1}$ we can discard the summation over $n$ entering the Eq.~\eqref{eq: hamiltonian s+p-wave chain} (i.e., only take the $n=0$ term) and find the leading behavior of the quantities entering $\mathcal{H}_{\mathsf{k}_1;s_2}^{\mathrm{chain}}$
as (note that $f_{\mathsf{k}_1;s_2}$ is manifestly positive)
\begin{sequation}
\begin{split}
\Delta_{\mathrm{s}}f_{\mathsf{k}_1;s_2}
+
\Delta_{\mathrm{t}}\tilde{g}_{\mathsf{k}_1;s_2}
\approx&\,
4m^2
\sum_{\lambda=\pm}
\int\frac{\mathrm{d}k_2}{2\pi}
\left(
\Delta_{\mathrm{s}}
\frac{1+s_2\lambda k_1/|\bs{k}|}{2\bs{k}^4}
+
\Delta_{\mathrm{t}}
\frac{\lambda|\bs{k}|+s_2 k_1}{2\bs{k}^4}
\right)
\\
=&\,
\frac{m^2}{|k_1|^3}
\left(\Delta_{\mathrm{s}}+s_2\, k_1\Delta_{\mathrm{t}}\right)
\end{split}
\end{sequation}
and 
\begin{sequation}
\begin{split}
g_{\mathsf{k}_1;s_2}
\approx&\,
2m\sum_{\lambda=\pm}
\int\frac{\mathrm{d}k_2}{2\pi}
\frac{1+s_2\lambda k_1/|\bs{k}|}{2\bs{k}^2}
\\
=&\,\frac{1}{|k_1|}.
\end{split}
\end{sequation}
Note that the first function is odd while the second function is even in $k_1$ to leading order for large $|k_1|$ as long as $\Delta_{\mathrm{t}}\neq0$.
As a consequence, the term multiplying $\tau_1$ changes sign at some $\mathsf{k}_1$ in the limit $\mathsf{a}\ll k^{-1}_{\pm}$ and for the function to be periodic is has to change sign twice. Let us denote the zeros of this term by $\mathsf{k}_{1,0}$ and $\mathsf{k}^\prime_{1,0}$ (in general there could be any even number of zeros, but we shall focus on the simplest case of two zeros here). The function $g_{\mathsf{k}_1;s_2}$ entering the Hamiltonian in the term proportional to $\tau_3$ will take two in general different values at $\mathsf{k}_{1,0}$ and $\mathsf{k}^\prime_{1,0}$, say $g_{\mathsf{k}_{1,0};s_2}<g_{\mathsf{k}^\prime_{1,0};s_2}$. Then, for impurity strengths $U$ such that
\begin{sequation}
\frac{2}{U}\in
\left(g_{\mathsf{k}_{1,0};s_2} , g_{\mathsf{k}^\prime_{1,0};s_2} \right)
\end{sequation}
the sign of the term proportional to $\tau_3$ changes sign between the two zeros of the term proportional to $\tau_1$. Hence, the winding number is nonzero and the chain is topological in this case.

\subsection{Scalar impurity chains on a chiral $p$-wave superconductor}
The Hamiltonian for a chiral $p$-wave superconductor
is readily obtained by evaluating Hamiltonian~\eqref{eq: hamiltonian s+p-wave chain} from the main text for one spin species (say $s_2=+1$) in the limit $\alpha\to0$ and $\Delta_{\mathrm{s}}\to0$.
In this case, $d_1(\mathsf{k}_1)$ defined in Eq.~\ref{eg: defintion d} is odd in $\mathsf{k}_1$ and therefore $d_1(0)=d_1(\pi/\mathsf{a})=0$, while $d_3(\mathsf{k}_1)$ is even in $\mathsf{k}_1$. We can thus conclude that the winding number $w$ is odd (and thus the phase topologically nontrivial), whenever $d_3(\mathsf{k}_1)$ changes sign between $\mathsf{k}_1=0$ and $\mathsf{k}_1=\pi/\mathsf{a}$, i.e., if
\begin{sequation}
\left(2/U-g_{0}\right)\left(2/U-g_{\pi/\mathsf{a}}\right)<0,
\label{eq: condition for w=1}
\end{sequation}
provided that the spectrum is fully gapped. 

In the limit of in infinitesimal $p$-wave gap $\Delta_{\mathrm{t}}k_{\mathrm{F}}\ll\mu$, 
the function $g$ from Eq.~\eqref{eq: f,g,tilde g} reduces to 
\begin{sequation}
\begin{split}
%f_{\mathsf{k}_1}:=&\,\frac{m}{k_{\mathrm{F}}\Delta_{\mathrm{t}}}
%\sum_{n}\frac{1}{\sqrt{k_{\mathrm{F}}^2-(\mathsf{k}_1+2\pi n)^2}},
%\\
g_{\mathsf{k}_1}:=&\,m
\sum_{n}\left[(\mathsf{k}_1+2\pi n/\mathsf{a})^2-k_{\mathrm{F}}^2\right]^{-1/2},
%\\
%\tilde{g}_{\mathsf{k}_1}:=&\,\frac{m}{k_{\mathrm{F}}\Delta_{\mathrm{t}}}
%\sum_{n}\frac{\mathsf{k}_1+2\pi n}{\sqrt{k_{\mathrm{F}}^2-(\mathsf{k}_1+2\pi n)^2}},
\end{split}
\label{eq: hamiltonian p-wave chain b}
\end{sequation}
where the sum is only taken over the values of $n$ for which the summand is real and $k_{\mathrm{F}}=\sqrt{2m\mu}$.
Observe that $g_{0}$ and $g_{\pi/\mathsf{a}}$ diverges when $k_{\mathrm{F}}\mathsf{a}/(2\pi)$ approaches from below an integer and half-integer value, respectively. We conclude that for sufficiently small impurity potentials $U$, there exists a series of topological phases for values of $\mathsf{a}$ slightly below half-integer or integer multiples of the Fermi wavelength $2\pi/k_{\mathrm{F}}$ (see Fig.~\ref{fig: phase diag} in the main text). The red and blue phase boundaries in \ref{fig: phase diag} are the values of $g_0/m$ and $g_{\pi/\mathsf{a}}/m$ defined in Eq.~\eqref{eq: hamiltonian p-wave chain b} and border the region where $w=1$ according to condition~\eqref{eq: condition for w=1}.

%\end{widetext}

\end{document}